\documentclass[12pt,preprint]{aastex}
\lefthead{K.S. Dawson \ea }
\righthead{Detection of Arcminute Anisotropy}

\slugcomment{Submitted to {\em Astrophysical Journal}}

\def\D{{\Delta}}

\def\kl{{\rm{k}\lambda}}
\def\ea{et al.\ }
\def\ah{^{\rm h}}
\def\am{^{\rm m}}
\def\as{^{\rm s}}
\def\pr{^{\prime}}
\def\2pr{^{\prime \prime}}

\def\greatsim{\mathrel{\raise.3ex\hbox{$>$\kern-.75em\lower1ex\hbox{$\sim$}}}}
\def\lesssim{\mathrel{\raise.3ex\hbox{$<$\kern-.75em\lower1ex\hbox{$\sim$}}}}
\begin{document}

\title{Measurement of Arcminute Scale Cosmic Microwave Background Anisotropy with 
the BIMA Array}

\author{
K.S.~Dawson\altaffilmark{1},
W.L.~Holzapfel\altaffilmark{1},
J.E.~Carlstrom\altaffilmark{2,3},\\
M.~Joy\altaffilmark{4},
S.J.~LaRoque\altaffilmark{2},
A.D.~Miller\altaffilmark{2,5}
and D.~Nagai\altaffilmark{2}}


\altaffiltext{1}{Department of Physics, University of California,
Berkeley CA 94720}
\altaffiltext{2}{Department of Astronomy and Astrophysics, University of Chicago, Chicago
IL 60637}
\altaffiltext{3}{Center for Cosmological Physics, Dept. of Physics, Enrico Fermi Institute,
University of Chicago, Chicago, IL 60637}
\altaffiltext{4}{Space Science Laboratory, SD50, NASA Marshall Space Flight Center,
Huntsville AL 35812}
\altaffiltext{5}{Hubble Fellow}
\authoremail{kdawson@cfpa.berkeley.edu}

\begin{abstract}
We report the results of our continued study of arcminute scale anisotropy in the Cosmic Microwave
Background (CMB) with the Berkeley-Illinois-Maryland Association (BIMA) array.
The survey consists of ten independent fields selected for low infrared
dust emission and lack of bright radio point sources.
With observations from the Very Large Array (VLA) at $4.8$ GHz, we have identified point sources which
could act as contaminants in estimates of the CMB power spectrum and removed them
in the analysis.
Modeling the observed power spectrum with a single flat band power with average
multipole of $\ell_{eff} = 6864$, we find
$\Delta T=14.2^{+4.8}_{-6.0}\,\mu$K at $68\%$ confidence.
The signal in the visibility data exceeds the expected contribution from
instrumental noise with $96.5\%$ confidence.
We have also divided the data into two bins corresponding to different spatial resolutions
in the power spectrum.
We find
$\Delta T_1=16.6^{+5.3}_{-5.9}\,\mu$K at $68\%$ confidence for CMB flat band power
described by an average multipole of $\ell_{eff} = 5237$ and
$\Delta T_2<26.5\,\mu$K at
$95\%$ confidence for $\ell_{eff} = 8748$.
\end{abstract}


\section{Introduction} \label{sec:intro}
Fluctuations in the distribution of matter at the epoch of recombination
create large angular scale anisotropy in the Cosmic Microwave Background (CMB).
This primordial anisotropy has been studied extensively
at degree and sub-degree angular scales in order to place constraints on the parameters
of cosmological models (Halverson et al. 2002, de Bernardis
et al. 2002, Scott et al. 2002, Lee et al. 2001, Miller \ea 2002, Padin et al. 2001).
At arcminute scales, the primordial anisotropy is damped to negligible amplitude
due to photon diffusion and the finite thickness of the last scattering surface (Hu \& White 1997).
On these smaller scales, secondary anisotropies
such as the Sunyaev-Zeldovich effect (SZE) are expected to dominate the
signal of the CMB power spectrum (Haiman $\&$ Knox 1999).
Studies of secondary anisotropy in the CMB 
have the potential to be a powerful probe of the growth of structure in the Universe.

In this paper, we report results from an ongoing program 
using the Berkeley-Illinois-Maryland Association (BIMA) interferometer 
to search for arcminute-scale 
CMB anisotropy.
Discussion of the instrument, data reduction,
expected signals (from both primary and secondary anisotropies) and 
previous measurements is included in earlier
publications (Holzapfel et al. 2000 and Dawson et al. 2001, hereafter H2000 and D2001 respectively). 
We describe observations and criteria for field selection in \S\ref{sec:obs}.
The Bayesian likelihood analysis used to constrain the CMB anisotropy is described in \S\ref{sec:anal}.
The results are presented 
in \S\ref{sec:results} including a discussion of tests for systematic errors in the analysis.
Finally, in \S\ref{sec:con}, we present the conclusion and comparison to simulations of the SZE.

\section{Observations} \label{sec:obs}
Analysis of eleven fields observed with the BIMA array during the summers of 1997, 1998,
and 2000 revealed a significant detection of excess power (D2001).
In an effort to achieve uniform sensitivity and selection criteria across our sample,
we continued the project in the summer of 2001 with a subset of eight fields from the original survey.
Two new fields were also added to the survey for a total of ten.
Each field was observed with the BIMA array and the Very Large Array (VLA)\footnote{The VLA is operated by
the National Radio Astronomy Observatory, a facility of the National
Science Foundation, operated under cooperative agreement
by Associated Universities, Inc.}.

\subsection{Field Selection} \label{sec:fields}
The two new fields added to the survey, BDF12 and BDF13, lie at 
Right Ascensions convenient for summer observations and outside of the galactic plane.
This makes for a total of ten independent fields
for the survey in 2001, covering approximately $0.1$ square degrees.
The fields were selected from the $IRAS \, 100\,\mu$m and VLA NVSS~\cite{NVSS}
radio surveys
to lie in regions with low dust contrast
and no bright radio sources.
In addition, we used SkyView to access
digitized sky survey and ROSAT images 
to check for bright optical or x-ray emission which could complicate 
follow-up observations.
The pointing centers for each of the ten fields 
are given in Table 1.

Two of the fields used in the D2001 analysis, PC1643+46 and VLA1312+32, were originally
selected to follow-up previously reported microwave decrements
(Jones et al. 1997; Richards et al. 1997).
A third field, in the direction of the high redshift quasar PSS0030+17, was originally
selected as a distant cluster candidate.
These three fields did not follow the same selection criteria as the rest of the fields
and are not included in the present sample.
Since observations of these fields are not as deep as those for the other blank fields,
removing them from the survey does not have a significant effect on the results reported in this paper.

\subsection{BIMA Observations}
A priority for 2001 observations was to achieve uniform flux sensitivity across the sample of
ten fields.
We re-observed fields from the previous years
and dedicated $55$ hours to each of the new fields, aiming for a uniform noise level of
$<150\,\mu$Jy/beam RMS on short baselines ($u$-$v < 1.1 \, \kl$) for each field in the sample.
This noise level corresponds to
$\sim 15 \,\mu$K RMS for a $2\pr$ synthesized beam.
An image of the field BDF13 observed with the BIMA array
is included in Figure 1 to illustrate the sensitivity in the $u$-$v$ range $0.63 - 1.1 \, \kl$.

All anisotropy observations were made using the BIMA array at Hat Creek.
Nine $6.1$ meter telescopes of the array
were equipped for operation at $28.5$ GHz, providing a $6.6\pr$ FWHM field of view.
In order to track phase and gain fluctuations, all field observations were
bracketed by observations of 
bright radio point sources (H2000).
The fluxes of these calibration sources are all referenced to the flux 
of Mars which is uncertain by approximately $4\%$ at $90\%$ confidence 
(see discussion in Grego 2001).
This uncertainty is small compared to the uncertainty in the
anisotropy signals we report here 
and therefore makes a negligible contribution to the uncertainties in the reported results.
The cumulative integration time for each of the ten fields observed in this survey
is listed in Table 1.
Figure 2 shows the result of imaging the field BDF13 using all of the $u$-$v$ data from the summer of 2001.
This image is a fair representation of the fields in the survey which typically have only $u$-$v$ data in the 
range $0.63-3.0 \, \kl$.
After calibration and data edits, a total of 607 hours of integration have been dedicated to this project.

\subsection{VLA Observations and Point Source Results} \label{sec:VLA}
Sources of foreground emission have the potential to contaminate estimates of the CMB power
spectrum.
Contaminants at $28.5$ GHz arising from galactic sources such as dust, synchrotron,
and free-free emission are
expected to be below the $\mu$K level at arcminute angular scales in the regions of the sky selected
for this survey. 
However, emission from radio point sources could contribute significantly
to excess power in the observations described here (e.g. Tegmark et al. 2000).

The compact configuration used for the BIMA anisotropy observations was designed to produce
high sensitivity to extended sources. 
However, this configuration has limited sensitivity on the long baselines which
in principle could be used
to locate and remove point sources.
To help constrain the contribution from point sources to the anisotropy 
measurements, we used the VLA
at a frequency of $4.8\,$GHz
to observe each field in the survey.
With $1.5$ hours per field, these observations yielded an RMS
flux of
$\sim 25\,\mu$Jy/beam for a $9\pr$ FWHM region with the same pointing center
as a BIMA field.
The positions of all point sources with flux $>6\sigma$ 
within $8\pr$ of the pointing center have been recorded.
Measured point sources with fluxes corrected for attenuation from the primary beam at $4.8$ GHz are listed in Table 2.
An image of the field BDF13 is shown in Figure 3
as an example of observations used to identify point sources
from the VLA data.

If the spectra of the point sources are relatively flat or falling, deep observations
with the VLA will identify those that lie near the noise level in the
$28.5$ GHz maps.
However, it is possible that a radio source with a steeply inverted spectrum may lie below the VLA detection threshold
but still contribute to the observations at $28.5$ GHz.
Advection dominated accretion flows are thought to be the most common inverted spectrum sources.
They typically have a slowly rising
spectrum, with a spectral index of $0.3$ to $0.4$ (Perna and DiMatteo, 2000).
Such a shallow spectrum would only provide a factor of two increase in flux between $4.8$ GHz and
$28.5$ GHz.

\section{Analysis} \label{sec:anal}
The analysis in this paper is similar to that used to produce the previous BIMA anisotropy results
and is based on the formalism presented in White et al. (1999) for the determination 
of the CMB power spectrum from interferometer data.
We bin the visibility data and calculate joint confidence intervals as described in H2000.
The likelihood function to test a theory for a set of bandpowers, $\{C_{\ell}\}$, with $n$ measured visibilities is
defined
\begin{equation}
{\cal L}\left( \{C_{\ell} \} \right) = {1\over\pi^n \det{C}} \ \exp
  \left[ -V^{*}({\bf u}_i) C_{ij}^{-1}V({\bf u}_j)\right] \quad,
\label{eqn:likelihood}
\end{equation}
where $C_{ij}$ is the correlation matrix of visibilities
at ${\bf u}_i$ and ${\bf u}_j$.

There are several changes from the analysis in H2000.
We first perform the analysis on a single bandpower with data spanning the $u$-$v$ range
$0.63-1.7 \, \kl$.
We then perform an analysis with data divided into two bins, $0.63-1.1 \, \kl$ and $1.1-1.7 \, \kl$, corresponding
to different spatial resolution. 
A constraint correlation matrix is introduced to account for the point sources identified with the VLA.
In this new formalism, the correlation matrix can be represented as
\begin{equation}
C_{ij}=C^{V}_{ij}+C^N_{ij}+C^{C}_{ij} \quad ,
\end{equation}
where $C^{V}_{ij}$ represents the theory correlation matrix, $C^N_{ij}$ represents the noise
correlation matrix, and $C^{C}_{ij}$ represents the constraint correlation matrix used to remove
the effect of point sources from the determination of the CMB power spectrum.

\subsection{Theory Correlation Matrix} \label{sec:ctheory}
The fundamental tool for analysis of Gaussian temperature fluctuations
is the theory correlation matrix.
The theory correlation matrix is calculated from the observed visibilities
measured at a set of points
${\bf u}_i$.
The measured flux densities are given by
\begin{equation}
V({\bf u}) = {\partial B_\nu\over\partial T} \int d{\bf x}
  \ {\Delta T({\bf x})}\ A({\bf x})
  e^{2\pi i{\bf u}\cdot{\bf x}} \quad ,
\end{equation}
where $\D T(\bf x)$ is the temperature distribution on the sky,
$A({\bf x})$ is the primary beam of the telescope,
\begin{equation}
{\partial B_\nu\over\partial T} =
  2k_B \left( {k_BT\over hc} \right)^2 {x^4 e^x\over (e^x-1)^2} \quad ,
\end{equation}
$k_B$ is Boltzmann's constant, and $x\equiv h\nu/k_{B}T_{\rm cmb}$.
Following White et al. (1999), we define the visibility
correlation matrix,
\begin{eqnarray}
C_{ij}^{V} & \equiv & \left\langle V^*({\bf u}_i)V({\bf u}_j)\right\rangle \nonumber\\
           &    =   & \left({\partial B_\nu\over\partial T}\right)^2
 \int_0^\infty w\,dw\; S(w) W_{ij}(w)\, ,
\end{eqnarray}
which is proportional to the product of the power spectrum, $S(w)$,
and the visibility window function.
The window function is given by
\begin{equation}
W_{ij}(|{\bf w}|) \equiv
    \int_0^{2\pi} d\theta_{w}\ \widetilde{A}^*({\bf u}_i-{\bf w})
    \widetilde{A}({\bf u}_j-{\bf w}) \quad ,
\label{eqn:wijdef}
\end{equation}
where ${\widetilde{A}({\bf u})}$ is the Fourier transform of the telescope
primary beam.
In the case of a single flat band power and $\ell > 60$, we can write
\begin{equation}
C_{ij}^{V} = {1\over{2 \pi}} \left({\partial B_\nu}\over{\partial T} \right)^2
\Delta T^2 \int_0^\infty\frac{dw}{w} W_{ij}(w) \quad ,
\end{equation}
where
\begin{equation}
\Delta T^2 \equiv {T_{CMB}^2\over {2 \pi}} C_{\ell}{\ell}{\left(\ell +1 \right)}
\end{equation}
is the RMS of the CMB anisotropy contributed by the power spectra $C_{\ell}$.
In general, the theory can be expressed as a function of bandpowers.
The most likely values and confidence intervals are calculated for each band independently
by integrating the likelihood function 
\begin{equation}
{\cal L}\left( \{C_{\ell m}\}\right) 
= \sum_{C_{\ell 1}} \cdots \sum_{C_{\ell m-1}} \sum_{C_{\ell m+1}} \cdots
\sum_{C_{\ell n}} {\cal L}\left( \{C_{\ell 1} \cdots C_{\ell n}\}\right) \quad .
\end{equation}

For the analysis of a single bin of visibility data in the $u$-$v$ range $0.63-1.7\,\kl$,
the likelihood is only calculated over one variable,
as in H2000 and D2001.
For the analysis in this work, it is assumed that $\Delta T^2 \geq 0$ and
for visibility data in the $u$-$v$ range $1.7-3.0\,\kl$, $C_{ij}^{V} = 0$.

\subsection{Noise Correlation Matrix}
As with most interferometric observations, the system temperature of the receivers in the BIMA
array is continuously
monitored.
Each visibility is assigned an estimated uncertainty
\begin{equation}
\sigma_{raw,i}^2 = \left({2k_B T_{sys,i}} \over {\eta A}\right)^2 {1 \over {\Delta_\nu t}} \quad , 
\end{equation}
where $T_{sys,i}$ is the system temperature for the visibility $V({\bf u}_i)$, $\eta$ is the aperture efficiency,
$A$ is the area of the telescope, $\Delta_\nu$ is the effective
bandwidth, and $t$ is the integration time.
The values of $T_{sys,i}$, $\eta$, and $\Delta_\nu$ are known within a few percent.
The noise is rescaled independently for each baseline.
The reduced $\chi^2$ is calculated from each visibility, j, recorded from a given baseline 
\begin{equation}
\chi^2 = \sum_j {V^*({\bf u}_j)V({\bf u}_j) \over \sigma_{raw,j}^2} \quad .
\end{equation}
The noise for that baseline is then rescaled so that $\sigma_j^2 = {\sigma_{raw,j}^2 \over \chi^2}$.
The noise correlation matrix is calculated from the rescaled noise for each baseline as in H2000,
$C^N_{ii}=\frac{1}{\sigma_i^2}$.
With approximately 3000 visibilities per baseline, the expected error in the rescaled noise due to sample variance
is $\sim 1.8\%$ on a single baseline.
In order to test what effect this uncertainty has on the analysis, we both increased and decreased each term in the
noise correlation matrix by $2\%$ and found no significant change in estimates of excess power.

\subsection{Constraint Correlation Matrix}
As discussed in D2001, the effectiveness of subtracting point sources directly from the visibility
data is limited with the compact
configuration of the BIMA array.
Using the VLA at $4.8$ GHz, we can identify the positions of point sources
which might contaminate the visibility
data.
We introduce a constraint correlation matrix (Bond et al. 1998)  to marginalize over point source fluxes
as in Halverson et al. (2002).
Using this formalism, it is necessary only to identify the position of a potential source
and not its flux.

Each point source is represented with a template
$\Lambda_{ni}$ of unknown amplitude $\kappa_n$ for each visibility $V({\bf u}_i)$ where the subscript,
$n$, corresponds to the $n$th point source.
The template is derived from the response of the visibilities to a point source that is offset from the pointing
center by $\Delta RA_n,\,\Delta DEC_n$ for
a baseline, $({\bf u}_i,{\bf v}_i)$
\begin{equation}
\Lambda_{ni}=e^{2\pi i (u_i \Delta RA_n + v_i \Delta DEC_n)} \quad .
\end{equation}
The constraint correlation matrix can be expressed 
\begin{equation}
C^{C}_{ij}= \sum_{n,n'} \Lambda_{ni}\kappa_n \Lambda_{n'j} \kappa_{n'} \quad ,
\end{equation}
where $n$ and $n'$ are each summed over the number of point sources in the field.
In the limit $\kappa_n \kappa_{n'} \rightarrow \infty$, this procedure effectively assigns zero weight
to modes within the visibility data that correspond to point source positions identified
with the VLA.
To avoid singularities in the inverted correlation matrix, a finite, but large
($\kappa_n \kappa_{n'} \gg \, \langle C^N_{ii} \rangle$) value is assigned to the
point source amplitude.
A value of $\kappa_n \kappa_{n'} \sim 10^4 \, \langle C^N_{ii} \rangle$ has been assumed in the analysis;
the results are insensitive to the exact value of $\kappa_n \kappa_{n'}$.

Although some fields were observed in 1998 with longer baselines, most
fields in this survey were observed from the compact configuration.
The compact configuration of the BIMA array was contained within a
$u$-$v$ radius of $3.0 \,\kl$.
These baselines are included in order to achieve a reasonable level of discrimination
between the signature of a point source and the CMB anisotropy described in \S\ref{sec:ctheory}.
In this manner, we remove the contribution to the measured power from 
linear combinations of baselines that may be corrupted by point source emission.
With each additional point source constraint, a degree of freedom is lost and the uncertainty in the
measured power increases accordingly.

\section{Results} \label{sec:results}
We have produced and analyzed images for each of the observed fields.
The statistics of the images are described in Table 3 ($0.63-1.7\,\kl$), 4 ($0.63-1.1\,\kl$), and 5
($1.1-1.7\,\kl$).
The reported RMS values are 
those expected from the noise properties of the visibilities.
The RMS temperature measurement corresponds to the synthesized beamsize for the given $uv$ coverage.
The baselines used to produce the image statistics
are the same as the baselines included in the two bins of
visibility data for the CMB anisotropy analysis.
The window function produced from the 
noise weighted sum of the window functions for the individual visibilities in the $u$-$v$ range $0.63-1.1\,\kl$
has an average value of $\ell_{eff}=5237$ with FWHM $\ell=2870$.
The window function for visibilities in the $u$-$v$ range $1.1-1.7\,\kl$ has an average value $\ell_{eff}=8748$
with FWHM $\ell=4150$.
For the analysis using a single bin of data in the $u$-$v$ range $0.63-1.7\,\kl$, 
$\ell_{eff}=6864$ with FWHM $\ell=6800$.
The window functions for all three bands are plotted as a function of multipole in Figure 4.

\subsection{Measurement of Anisotropy}
We present the results of the analysis of the BIMA data assuming a CMB power spectrum that is described
by one or two flat band powers.
In Tables 3, 4, and 5,
we show the most likely $\Delta T$ with confidence intervals for each of the ten fields
included in this survey.
Results for the
joint likelihood for all the data are also included.
Figure~5 shows the relative likelihoods as a function
of assumed $\Delta T$ for a combined likelihood
analysis of the ten fields.
The results are normalized to unity likelihood for the case of no 
anisotropy signal.
The measured signal exceeds that expected from instrument noise with $97.8\%$ confidence for
data which falls in the $u$-$v$ range $0.63-1.1\,\kl$ and with $96.5\%$ confidence for
the single bin of visibility data that covers the $u$-$v$ range $0.63-1.7\,\kl$.

We include a contour plot of the two dimensional likelihood function in Figure~6 in order
to demonstrate the correlation between the two flat band power bins.
Contours represent $68\%$, $95\%$, and $99\%$ confidence intervals.
We calculate the correlation between the two bandpower bins directly from the 
likelihood function,
\begin{equation}
F =
\left(
\begin{array}{cc}
\langle ( \Delta T_1-\overline{\Delta T}_{1})^2 \rangle & \langle(\Delta T_1-\overline{\Delta T}_{1})(\Delta T_2-\overline{\Delta T}_{2})  \rangle \\
\langle (\Delta T_2-\overline{\Delta T}_{2})(\Delta T_1-\overline{\Delta T}_{1})\rangle & \langle (\Delta T_2-\overline{\Delta T}_{2})^2 \rangle
\end{array}
\right) 
\end{equation}

\begin{equation}
F = \\
\left(
\begin{array}{cc}
3.738 \times 10^{-11} & 1.448 \times 10^{-11} \\
1.448 \times 10^{-11} & 1.920 \times 10^{-10}
\end{array}
\right)
\,(\mu K)^2 \quad .
\end{equation}

This matrix can be diagonalized with eigenvalues
\begin{equation}
F\pr = \\
\left(
\begin{array}{cc}
3.604 \times 10^{-11} & 0 \\
0 & 1.933 \times 10^{-10}
\end{array}
\right)
\,(\mu K)^2 \quad ,
\end{equation}
and corresponding eigenvectors, ${\bf X}_1\pr = 0.996{\bf X}_1 + 0.0893{\bf X}_2$ and 
${\bf X}_2\pr = 0.0893{\bf X}_1 - 0.996{\bf X}_2$.
There is less than $10\%$ correlation between the two bins of visibility data.

\subsection{Systematics Check} \label{sec:systematics}

We performed tests for systematic errors in all fields with significant excess power as described
in D2001.  
Due to finite computing resources, the analysis is limited to visibility data in $u$-$v$ range $0.63-1.1\,\kl$
described by $\Delta T_1$.
The power in the second bin is fixed at
$\Delta T_2=0$ for all tests described
in this section.
We tested for sources of contamination that change with time.
Assuming that a terrestrial source, the sun, or the moon varies in position by several degrees over the
course of a few days, such a local effect should be discovered from analysis of
subsets divided into hours of observation
or days of observation described in D2001.
For the field BDF12, an analysis of three subsets divided into time of observation and three subsets of
date of observation revealed a range of values $\Delta T_1 = 16.2^{+21.4}_{-16.2} \, \mu$K to
$\Delta T_1 = 71.0^{+33.5}_{-28.0} \, \mu$K, in each case consistent with the reported value of
$\Delta T_1 = 44.2^{+16.4}_{-13.7} \, \mu$K.
For the field BDF13, the various subsets were found to have most likely values in the range
$\Delta T_1 = 33.8 \, \mu$K to $\Delta T_1 = 58.9 \, \mu$K, consistent with the reported
value of $\Delta T_1 = 38.8^{+17.2}_{-15.9} \, \mu$K.
There was no significant difference in the measured power in subsets divided
by day of observation or in the subsets divided by hour of
observation.

We also added three additional tests this year to search for systematic
errors in BDF12 and BDF13.
We performed a jack-knife systematic test by breaking the
data into subsets of four and five telescopes looking for antenna based
systematic errors.
For a test of baseline based systematic errors, we created an east-west baseline subset and
a north-south baseline subset.
Subsets for BDF12 had most likely values in the range
$\Delta T_1 = 30.7 \, \mu$K to $\Delta T_1 = 53.0 \, \mu$K while subsets for BDF13 had most likely
values in the range $\Delta T_1 = 37.2 \, \mu$K to $\Delta T_1 = 41.9 \, \mu$K.
There was approximately the same level of excess power in each subset, as well as all of the
subsets described in D2001.
The third test combined the raw visibilities from observations of several different fields taken in
a single summer to look for correlations between observations of independent fields.
As would be expected for uncorrelated data, the analysis of the combined
data sets revealed a measurement of excess power that is consistent with instrumental noise
to $68\%$ confidence.
Overall, we find no evidence that our results are biased by systematic effects.

As discussed in \S\ref{sec:VLA}, we have adopted a VLA flux density limit of $6\sigma$ for identifying point sources.
We now consider how different point source flux density limits affect our estimates of
CMB anisotropy.
We test the effect of four different point source models on the determination of the most likely
$\Delta T_1$ and confidence intervals.
The results are found in Table 6.
There is no significant difference in the $\Delta T_1$ determined from the range of models considered.
As in D2001, the flat band power results are insensitive to the details of the point source
analysis.

\section{Conclusion} \label{sec:con}
Over the course of three summers, we have used the BIMA array in a compact 
configuration at $28.5\,$GHz to search for CMB anisotropy in ten
independent $6.6^\prime$ FWHM fields.
With these observations, we have detected arcminute scale
anisotropy at better than $95\%$ confidence.
In the context of an assumed flat band power model for the CMB power spectrum,
we find $\Delta T=14.2^{+4.8}_{-6.0}\,\mu$K at $68.3\%$ confidence
with sensitivity on scales that
correspond to an average harmonic multipole $\ell_{eff} = 6864$.
We also present results after dividing the visibility data into two bins of
different spatial resolution.
We find $\Delta T_1=16.6^{+5.3}_{-5.9}\,\mu$K at $68.3\%$ confidence 
on scales corresponding to an average harmonic multipole $\ell_{eff} = 5237$
and $\Delta T_2 < 26.5 \,\mu$K at $95\%$ confidence 
at $\ell_{eff} = 8748$.
The results of the VLA observations appear to exclude the possibility of point source contamination and there is
no indication of an obvious systematic error that would bias the observations.

Mason et al. (2002) have also reported a detection of excess power at somewhat larger angular scales.
They find $\Delta T=22.5^{+2.5}_{-3.6}\,\mu$K for data in the range $2010 \, < \, \ell \, < 4000$.
Although this measurement is at a lower $\ell$ than the BIMA results, it
is significantly higher than the expected power due to primordial anisotropy.

If the signal observed in the BIMA fields is indeed caused by CMB anisotropy, there are a host of possible sources for
excess power such as primary anisotropy, thermal SZE, kinetic SZE, patchy
reionization, and the Ostriker-Vishniac effect.
Of these candidates for CMB anisotropy, the thermal SZE from clusters of galaxies is expected to dominate on
the scales where the BIMA instrument is most sensitive (see for example, Gnedin \& Jaffe 2001).
Analytic models and simulations of cluster formation predict $\Delta T$ values that range from
$4.3\,\mu$K to $15.0\,\mu$K on angular
scales of approximately two arcminutes.
Figure 7 compares the results of this paper to the theoretical models. 

The non-Gaussian characteristics of the CMB power spectrum caused by the thermal SZE
may increase the uncertainty in measurements of the power spectrum due to sample variance.
Current models suggest that these effects increase the uncertainty by a factor of 3 over what is expected for
the sample variance of a Gaussian distributed signal at $\ell \sim 5000$ (White, Hernquist, \& Springel 2002).
Based on this argument, the effect of non-Gaussian sample variance contributes a standard deviation
of $3 \times \sqrt{2/N} \,  \Delta T = 6.0 \, \mu$K to the anisotropy measurements, where $N\sim 100$ is the
number of independent pixels in the $u$-$v$ range $0.63-1.1\,\kl$ and $\sqrt{2/N} \, \Delta T$ is the sample variance in
a Gaussian distributed signal.
The uncertainty due to sample variance is approximately equal to the statistical uncertainty reported
in this paper, increasing the overall uncertainty by $40 \%$, in agreement with the predictions of
Zhang,
Pen $\&$ Wang (2002).
The predicted effect of sample variance on the measurements is represented by the extended error bars
in Figure 7.

\vskip 20 pt

We thank the entire staff of the BIMA observatory for their
many contributions to this project, in particular
Rick Forster and Dick Plambeck for their assistance with both the 
instrumentation and observations.
Nils Halverson and Martin White are thanked for stimulating discussions concerning
data analysis. 
We are grateful
for the scheduling of time at the VLA in support of this project that has proved essential
to the point source treatment.
This work is supported in part by NASA LTSA grant number NAG5-7986,
NSF grant 0096913, and the David and Lucile Packard Foundation.
The BIMA millimeter array is supported by NSF grant AST 96-13998. 
A.M. is supported by Hubble Grant ASTR/HST-HF-0113.

\newpage
\markright{REFERENCES}

\newpage
\markright{TABLES}

\begin{table*}[htb]
\begin{center}
\begin{tabular}{lcccc}
\multicolumn{5}{c}{TABLE 1}\\
\multicolumn{5}{c}{Field Positions and Observation Times}\\\hline\hline
\multicolumn{1}{c}{Fields} & R. A. (J2000) & Decl. (J2000)  & Observation year(s) & Time (Hrs) \\ \hline
BDF4 & $00\ah\,28\am\,04.4\as$ & $+28^{\circ}\,23\pr\,06\2pr$ & 98 &  $77.6$\\
HDF & $12\ah\,36\am\,49.4\as$ & $+62^{\circ}\,12\pr\,58\2pr$ & 98, 01 &  $59.9$\\
BDF6 & $18\ah\,21\am\,00.0\as$ & $+59^{\circ}\,15\pr\,00\2pr$ & 98, 00 &  $81.2$\\
BDF7 & $06\ah\,58\am\,45.0\as$ & $+55^{\circ}\,17\pr\,00\2pr$ & 98, 00 & $68.2$\\
BDF8 & $00\ah\,17\am\,30.0\as$ & $+29^{\circ}\,00\pr\,00\2pr$ & 00, 01 & $53.3$\\
BDF9 & $12\ah\,50\am\,15.0\as$ & $+56^{\circ}\,52\pr\,30\2pr$ & 00, 01 &  $53.9$\\
BDF10 & $18\ah\,12\am\,37.2\as$ & $+58^{\circ}\,32\pr\,00\2pr$ & 00, 01 & $53.3$\\
BDF11 & $06\ah\,58\am\,00.0\as$ & $+54^{\circ}\,24\pr\,00\2pr$ & 00, 01 & $50.0$\\
BDF12 & $06\ah\,57\am\,38.0\as$ & $+55^{\circ}\,32\pr\,00\2pr$ & 01 &  $54.8$\\
BDF13 & $22\ah\,22\am\,45.0\as$ & $+36^{\circ}\,37\pr\,00\2pr$ & 01 &  $54.5$\\
\end{tabular}
\end{center}
\label{tab:coord}
\end{table*}

\begin{table*}[htb]
\begin{center}
\begin{tabular}{lccc}
\multicolumn{4}{c}{TABLE 2}\\
\multicolumn{4}{c}{Point Sources From VLA at $4.8$ GHz}\\\hline\hline
\multicolumn{1}{c}{Field} &  $\Delta$ R. A. ($^{\2pr}$) & $\Delta$ Decl. ($^{\2pr}$) & Intrinsic Flux ($\mu$Jy) \\ \hline
BDF4 & $-96.8$ & $\phn 255.7$ & $1230$ \\
BDF4 & $\phn \phn 72.8$ & $\phn 178.2$ & $\phn 514$ \\
BDF4 & $\phn \phn 99.9$ & $-89.4$ & $\phn 221$ \\
BDF4 & $-94.9$ & $\phn 268.4$ & $\phn 391$ \\
BDF4 & $-121.0$ & $\phn 356.3$ & $\phn 673$ \\
HDF & $-35.0$ & $-85.0$ & $\phn 591$ \\
HDF & $\phn 255.0$ & $-89.8$ & $\phn 998$ \\
HDF & $\phn 178.3$ & $-274.0$ & $\phn 731$ \\
HDF & $\phn 222.5$ & $-86.8$ & $\phn 345$ \\
HDF & $\phn \phn 69.1$ & $\phn 334.1$ & $\phn 593$ \\
HDF & $-21.1$ & $\phn \phn 66.1$ & $\phn 160$ \\
BDF6 & $-136.5$ & $-283.5$ & $\phn 592$ \\
BDF6 & $-247.2$ & $\phn 259.9$ & $\phn 510$ \\
BDF7 & $\phn 314.6$ & $\phn \phn 47.4$ & $1554$ \\
BDF7 & $\phn 173.8$ & $\phn \phn 97.8$ & $\phn 373$ \\
BDF7 & $\phn 253.8$ & $\phn -1.1$ & $\phn 284$ \\
BDF8 & $-145.9$ & $-266.1$ & $1381$ \\
BDF8 & $\phn \phn 27.6$ & $\phn 280.9$ & $\phn 611$ \\
BDF8 & $\phn 302.9$ & $-79.5$ & $\phn 622$ \\
BDF9 & $-221.9$ & $-123.7$ & $1500$ \\
BDF9 & $-192.7$ & $\phn 215.8$ & $1193$ \\
BDF9 & $\phn 245.2$ & $-101.0$ & $1039$ \\
BDF9 & $\phn 438.3$ & $-147.3$ & $2979$ \\
BDF10 & $-158.5$ & $-165.6$ & $1670$ \\
BDF10 & $-146.1$ & $-183.9$ & $\phn 320$ \\
\end{tabular}
\end{center}
\label{tab:vla1}
\end{table*}

\begin{table*}[htb]
\begin{center}
\begin{tabular}{lccc}
\multicolumn{4}{c}{TABLE 2 cont'd}\\
\multicolumn{4}{c}{Point Sources From VLA at $4.8$ GHz}\\\hline\hline
\multicolumn{1}{c}{Field} &  $\Delta$ R. A. ($^{\2pr}$) & $\Delta$ Decl. ($^{\2pr}$) & Intrinsic Flux ($\mu$Jy) \\ \hline
BDF11 & $\phn \phn 87.7$ & $\phn \phn 77.9$ & $\phn 246$ \\
BDF11 & $\phn 342.8$ & $\phn \phn \phn 8.8$ & $\phn 536$ \\
BDF11 & $\phn \phn 42.5$ & $-11.8$ & $\phn 152$ \\
BDF12 & $-398.3$ & $-115.1$ & $2796$ \\
BDF12 & $-241.0$ & $-256.7$ & $1620$ \\
BDF12 & $\phn 260.2$ & $\phn 300.5$ & $1191$ \\
BDF12 & $-382.7$ & $-125.9$ & $\phn 937$ \\
BDF12 & $-137.4$ & $-133.4$ & $\phn 278$ \\
BDF12 & $\phn 170.9$ & $\phn \phn 66.7$ & $\phn 211$ \\ 
BDF13 & $-78.9$ & $\phn 431.7$ & $6244$ \\
BDF13 & $\phn 181.4$ & $-49.5$ & $\phn 477$ \\
BDF13 & $-154.0$ & $\phn 299.7$ & $1145$ \\
BDF13 & $\phn 225.1$ & $-99.9$ & $\phn 317$ \\
\end{tabular}
\end{center}
\label{tab:vla2}
\end{table*}

\begin{table*}[htb]
\begin{center}
\begin{tabular}{lcccccc}
\multicolumn{7}{c}{TABLE 3}\\
\multicolumn{7}{c}{Image Statistics, Most Likely $\Delta T$, and Confidence Intervals for
$0.63-1.7\,\kl$}\\\hline\hline
\multicolumn{1}{l}{} & Synthesized & RMS & RMS & \multicolumn{3}{c}{$\Delta T(\mu{\rm K})$} \\
\multicolumn{1}{c}{Field} &  Beamsize($^{\2pr}$) & {($\mu$Jy$\,$beam$^{-1}$)} & ($\mu$K)
& Most Likely & $68\%$ & $95\% $ \\\hline
BDF4 &  $87.1 \times 94.6$ & $103.5$ & $18.8$ & $\phn 0.0$ & $\phn 0.0-14.4$ & $0.0-29.4$ \\
HDF &  $87.4 \times 90.9$ & $111.3$ & $21.0$ & $\phn 0.0$ & $\phn 0.0-17.4$ & $0.0-34.0$ \\
BDF6 &  $86.6 \times 90.9$ & $\phn 89.9$ & $17.3$ & $24.0$ & $14.2-34.6$ & $2.4-44.6$ \\
BDF7 &  $86.4 \times 90.4$ & $101.6$ & $19.5$ & $17.4$ & $\phn 2.6-27.6$ & $0.0-44.2$ \\
BDF8 &  $84.1 \times 87.5$ & $108.5$ & $22.1$ & $\phn 0.0$ & $\phn 0.0-12.6$ & $0.0-24.8$ \\
BDF9 &  $85.3 \times 88.6$ & $110.6$ & $22.0$ & $12.4$ & $\phn 0.0-21.6$ & $0.0-37.6$ \\
BDF10 &  $85.8 \times 86.8$ & $108.6$ & $21.9$ & $\phn 0.0$ & $\phn 0.0-15.6$ & $0.0-30.0$ \\
BDF11 &  $85.2\times 88.3$ & $109.1$ & $21.8$ & $\phn 0.0$ & $\phn 0.0-17.6$ & $0.0-33.6$ \\
BDF12 &  $87.3 \times 88.6$ & $112.3$ & $21.8$ & $35.8$ & $23.2-50.2$ & $9.8-67.2$ \\
BDF13 &  $86.8 \times 89.1$ & $112.9$ & $21.9$ & $27.2$ & $11.4-41.4$ & $0.0-53.0$ \\\\
All Fields & $$ & $$ & $$ & $14.2$ & $\phn 8.2-19.0$ & $0.8-21.8$ \\
\end{tabular}
\end{center}
\end{table*}

\begin{table*}[htb]
\begin{center}
\begin{tabular}{lcccccc}
\multicolumn{7}{c}{TABLE 4}\\
\multicolumn{7}{c}{Image Statistics, Most Likely $\Delta T$, and Confidence Intervals for
$0.63-1.1\,\kl$}\\\hline\hline
\multicolumn{1}{l}{} & Synthesized & RMS & RMS & \multicolumn{3}{c}{$\Delta T_1(\mu{\rm K})$} \\
\multicolumn{1}{c}{Field} &  Beamsize($^{\2pr}$) & {($\mu$Jy$\,$beam$^{-1}$)} & ($\mu$K)
& Most Likely & $68\%$ & $95\% $ \\\hline
BDF4 &  $119.7 \times 123.3$ & $141.7$ & $14.4$ & $\phn 0.0$ & $\phn 0.0-13.5$ & $\phn 0.0-28.2$ \\
HDF &  $114.2 \times 123.8$ & $151.8$ & $16.1$ & $\phn 0.0$ & $\phn 0.0-21.1$ & $\phn 0.0-41.1$ \\
BDF6 &  $118.2 \times 119.5$ & $122.3$ & $13.0$ & $21.1$ & $\phn 9.5-33.2$ & $\phn 0.0-44.3$ \\
BDF7 &  $117.7 \times 121.2$ & $138.6$ & $14.6$ & $23.4$ & $\phn 5.6-39.1$ & $\phn 0.0-57.4$ \\
BDF8 &  $108.5 \times 121.0$ & $151.1$ & $17.3$ & $\phn 0.0$ & $\phn 0.0-14.3$ & $\phn 0.0-28.5$ \\
BDF9 &  $110.6 \times 117.7$ & $149.7$ & $17.2$ & $15.2$ & $\phn 0.0-24.5$ & $\phn 0.0-42.9$ \\
BDF10 &  $111.0 \times 116.8$ & $148.8$ & $17.2$ & $\phn 0.0$ & $\phn 0.0-17.5$ & $\phn 0.0-33.8$ \\
BDF11 &  $110.9 \times 117.7$ & $149.1$ & $17.1$ & $\phn 0.0$ & $\phn 0.0-23.6$ & $\phn 0.0-44.8$ \\
BDF12 &  $110.3 \times 117.1$ & $149.2$ & $17.3$ & $44.2$ & $30.5-60.6$ & $20.9-76.3$ \\
BDF13 &  $110.8 \times 121.0$ & $154.2$ & $17.3$ & $38.8$ & $22.9-56.0$ & $\phn 7.1-71.8$ \\\\
All Fields & $$ & $$ & $$ & $16.6$ & $10.7-21.9$ & $\phn 2.5-26.0$ \\
\end{tabular}
\end{center}
\end{table*}

\begin{table*}[htb]
\begin{center}
\begin{tabular}{lcccccc}
\multicolumn{7}{c}{TABLE 5}\\
\multicolumn{7}{c}{Image Statistics, Most Likely $\Delta T$, and Confidence Intervals for
$1.1-1.7\,\kl$}\\\hline\hline
\multicolumn{1}{l}{} & Synthesized & RMS & RMS & \multicolumn{3}{c}{$\Delta T_2(\mu{\rm K})$} \\
\multicolumn{1}{c}{Field} &  Beamsize($^{\2pr}$) & {($\mu$Jy$\,$beam$^{-1}$)} & ($\mu$K)
& Most Likely & $68\%$ & $95\% $ \\\hline
BDF4 &  $69.5 \times 77.4$ & $153.2$ & $42.7$ &
$33.2$ & $5.3-49.6$ & $0.0-69.0$ \\ 
HDF &  $69.3 \times 75.7$ & $164.5$ & $47.0$ &
$26.5$ & $0.0-43.2$ & $0.0-68.4$ \\
BDF6 &  $69.5 \times 74.4$ & $132.7$ & $38.5$ &
$10.7$ & $0.0-32.9$ & $0.0-57.5$ \\
BDF7 &  $69.6 \times 73.7$ & $149.2$ & $43.6$ &
$\phn 2.2$ & $0.0-38.4$ & $0.0-65.3$ \\
BDF8 &  $67.6 \times 74.6$ & $155.8$ & $46.3$ &
$\phn 0.0$ & $0.0-29.6$ & $0.0-55.8$ \\
BDF9 &  $69.8 \times 72.0$ & $164.1$ & $49.0$ &
$\phn 0.0$ & $0.0-34.7$ & $0.0-61.8$ \\
BDF10 &  $70.6 \times 71.1$ & $159.0$ & $47.5$ &
$\phn 0.0$ & $0.0-30.7$ & $0.0-58.3$ \\
BDF11 &  $68.8 \times 73.7$ & $159.9$ & $47.3$ &
$\phn 0.0$ & $0.0-29.6$ & $0.0-56.9$ \\
BDF12 &  $70.3 \times 72.2$ & $170.7$ & $50.4$ & 
$\phn 0.0$ & $0.0-36.3$ & $0.0-64.2$ \\
BDF13 &  $69.6 \times 75.0$ & $165.7$ & $47.6$ & 
$\phn 0.0$ & $0.0-29.8$ & $0.0-57.4$ \\\\
All Fields & $$ & $$ & $$ & 
$\phn 0.0$ & $0.0-14.6$ & $0.0-26.5$ \\\\
\end{tabular}
\end{center}
\end{table*}

\begin{table*}[htb]
\begin{center}
\begin{tabular}{lcccc}
\multicolumn{5}{c}{TABLE 6}\\
\multicolumn{5}{c}{The Effect of point source model on $\Delta T_1$}\\\hline\hline
\multicolumn{1}{l}{} & \multicolumn{3}{c}{$\Delta T_1(\mu{\rm K})$} & Confidence\\
\multicolumn{1}{l}{VLA Flux Limit} & {Most likely} & $68\%$ & $95\%$ & $\Delta T_1>0$ \\ \hline
none & $17.7$ & $12.1-22.9$ & $4.7-27.6$ & $98.8\%$ \\
$>12\sigma$ & $16.7$ & $10.9-21.7$ & $2.8-26.0$ & $98.1\%$ \\
$>8\sigma$ & $17.1$ & $11.2-22.3$ & $3.7-26.7$ & $98.3\%$ \\
$>6\sigma$ & $16.4$ & $10.5-21.7$ & $3.1-25.7$ & $97.9\%$ \\
\end{tabular}
\end{center}
\label{tab:qflat}
\end{table*}

\clearpage
\markright{figures}

\begin{figure}
\figurenum{1}
\plotone{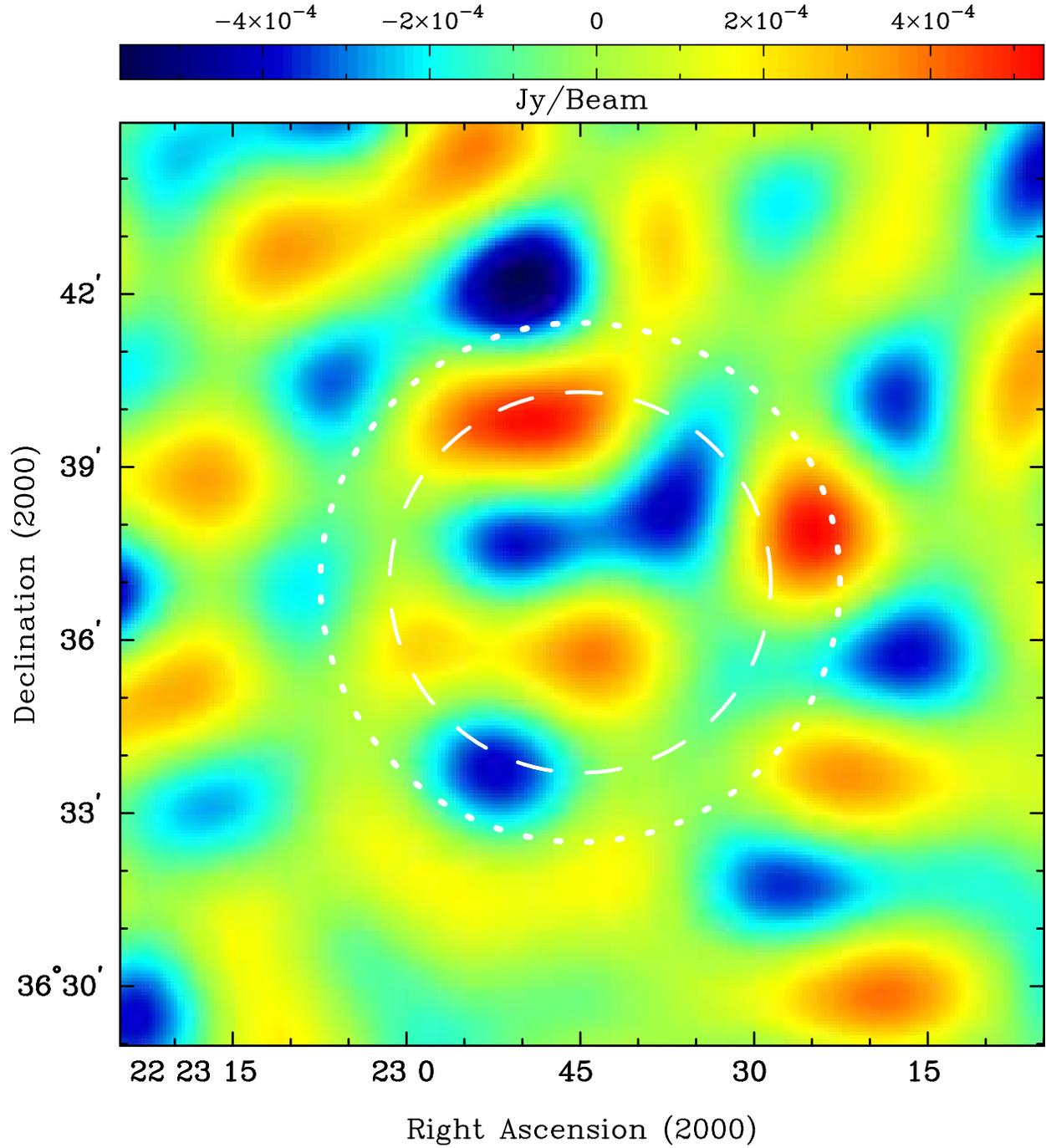}
\figcaption[f1.eps]{
BIMA image of the field BDF13 using only data in the $u$-$v$ range $0.63-1.1\,\kl$.
The corresponding image statistics can be found in Table 4.
The inner dashed line represents the $6.6\pr$ FWHM field of view of the BIMA instrument.
The outer dotted line represents the $9.0\pr$ FWHM field of view of the VLA.
}
\label{fig:bdf13_ran}
\end{figure}

\begin{figure}
\figurenum{2}
\plotone{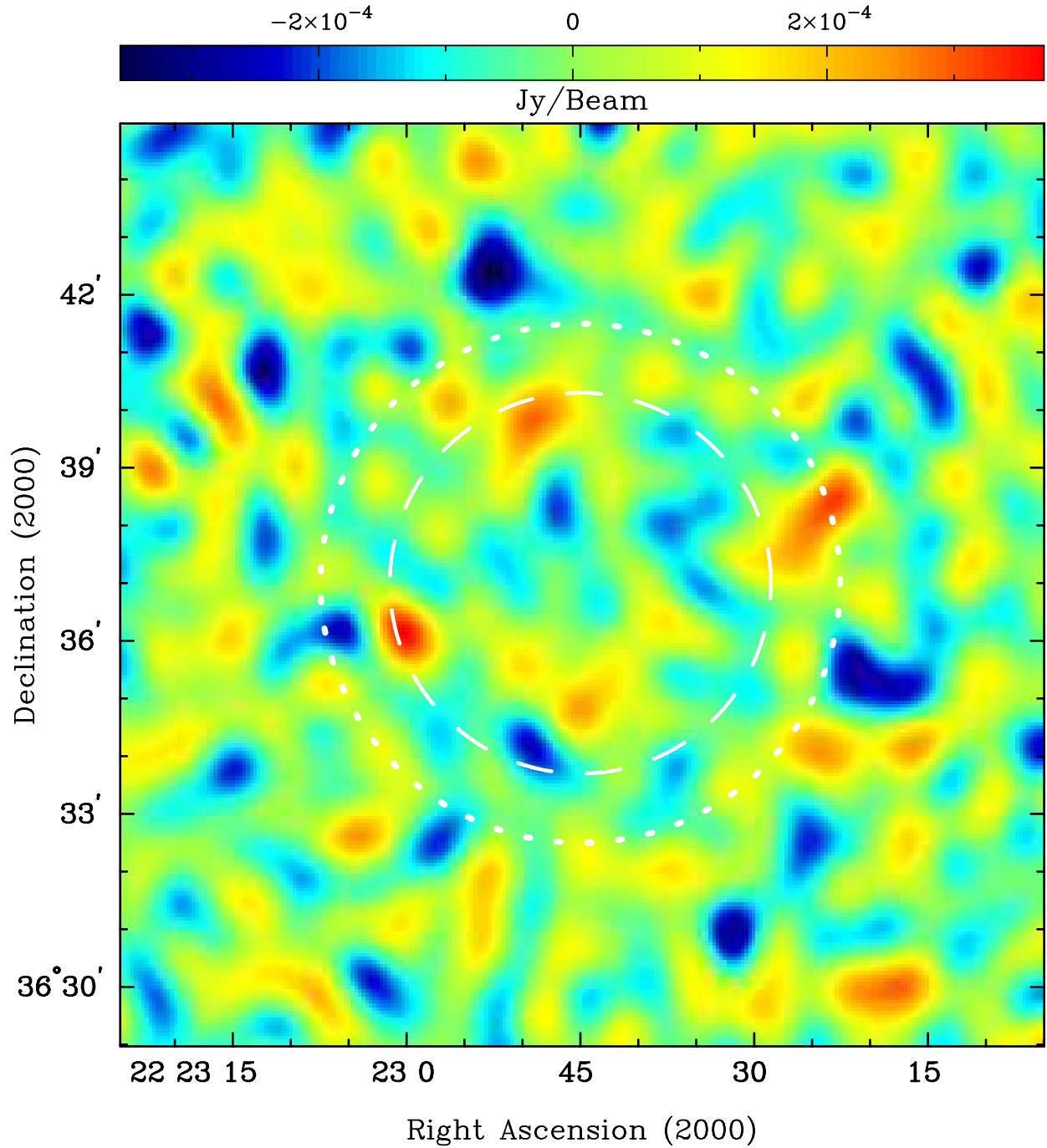}
\figcaption[f2.eps]{
BIMA image of the field BDF13 using all data taken in the summer of 2001.
The measured RMS in the image is $97.2 \, \mu$Jy/beam.
The synthesized beam is described by a Gaussian FWHM of $62.5 \2pr$ by $57.6 \2pr$.
}
\label{fig:bdf13_all}
\end{figure}

\begin{figure}
\figurenum{3}
\plotone{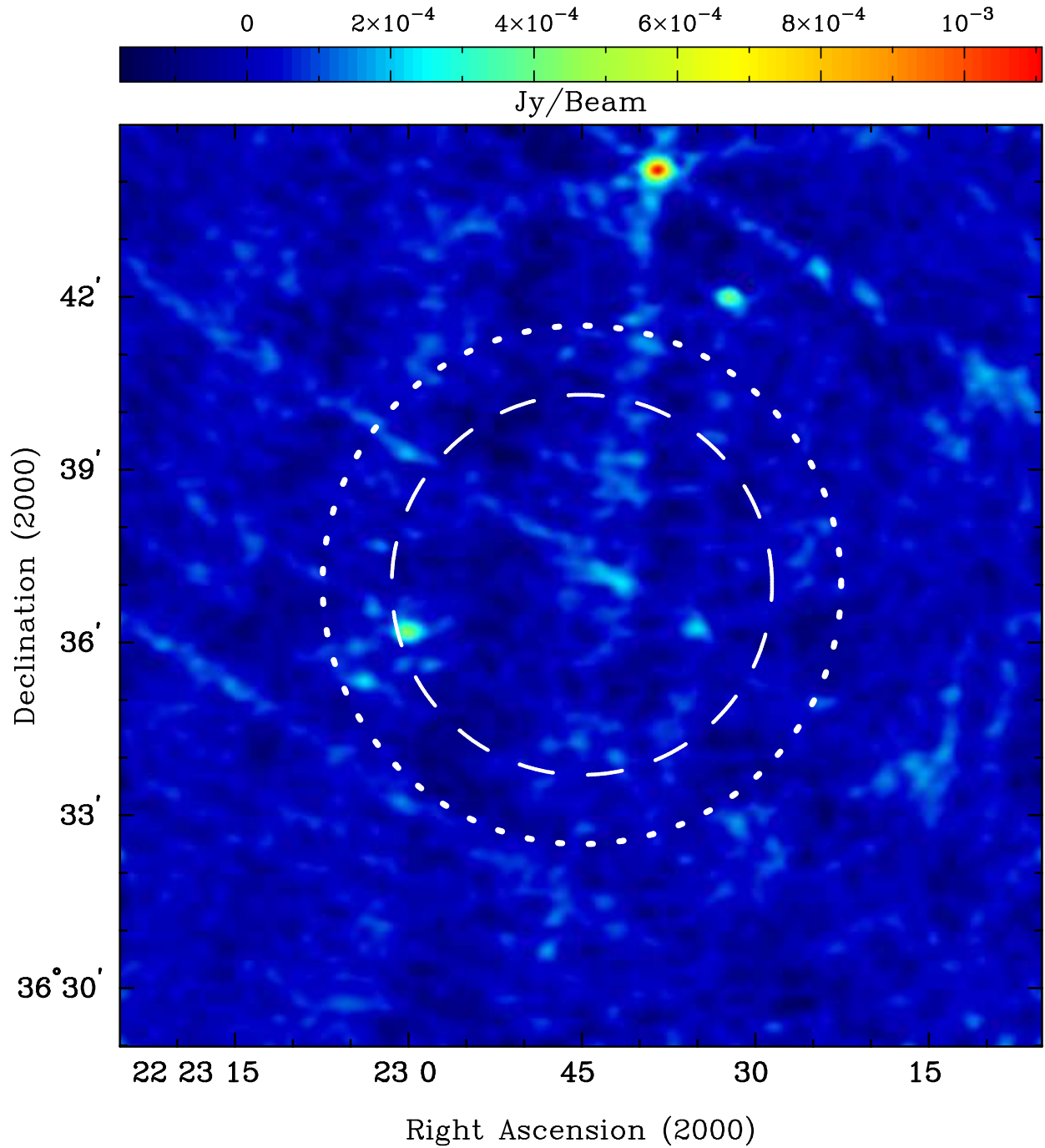}
\figcaption[f3.eps]{
VLA image of the field BDF13 used to identify radio point sources at $4.8$ GHz.
The measured RMS in the image is $28.6 \, \mu$Jy/beam after removal of the 
point sources listed in Table 2.
The synthesized beam is described by a Gaussian FWHM of $21.4 \2pr$ by $17.9 \2pr$.
}
\label{fig:vla}
\end{figure}

\begin{figure}
\figurenum{4}
\plotone{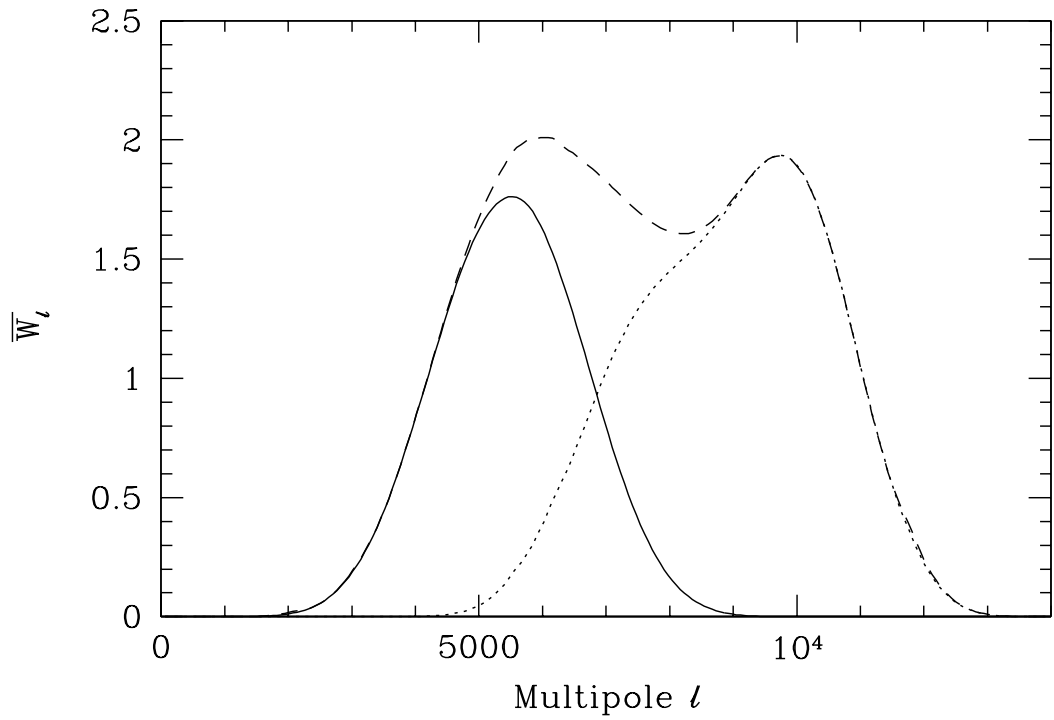}
\figcaption[f4.eps]{
The ``data weighted'' window functions $\overline{W}_{\ell}$.
The solid line corresponds to data in the $u$-$v$ range $0.63-1.1\,\kl$,
the dotted line corresponds to data in the $u$-$v$ range $1.1-1.7\,\kl$
and the dashed line corresponds to data in the $u$-$v$ range $0.63-1.7\,\kl$.
The dual peaks are due to the bimodal distribution of baselines in the
compact BIMA configuration.}
\label{fig:win}
\end{figure}

\begin{figure}
\figurenum{5}
\plotone{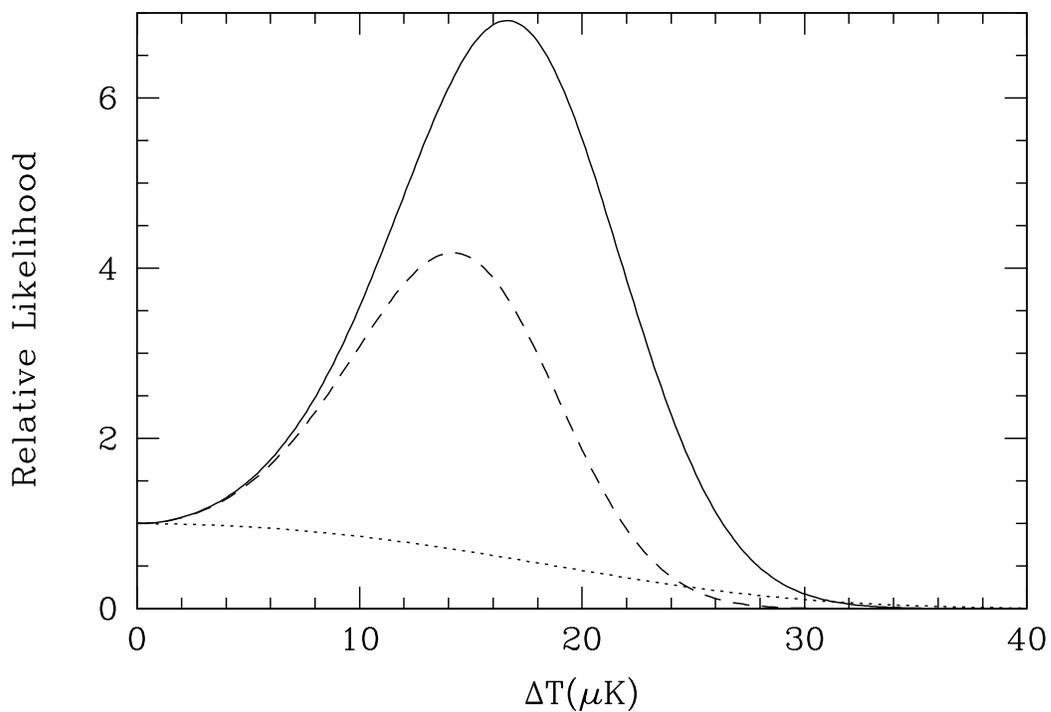}
\figcaption[f5.eps]
{The relative likelihood that the observed signal in the ten combined fields
is described by flat band power with amplitude $\Delta T$.
The solid line corresponds to an analysis of $\Delta T_1$, with data in the
$u$-$v$ range $0.63-1.1\,\kl$,
the dotted line corresponds to an analysis of $\Delta T_2$, with data in the $u$-$v$ range $1.1-1.7\,\kl$
and the dashed line corresponds to an analysis of $\Delta T$, with data in the $u$-$v$ range $0.63-1.7\,\kl$.}
\label{fig:all1}
\end{figure}

\begin{figure}
\figurenum{6}
\plotone{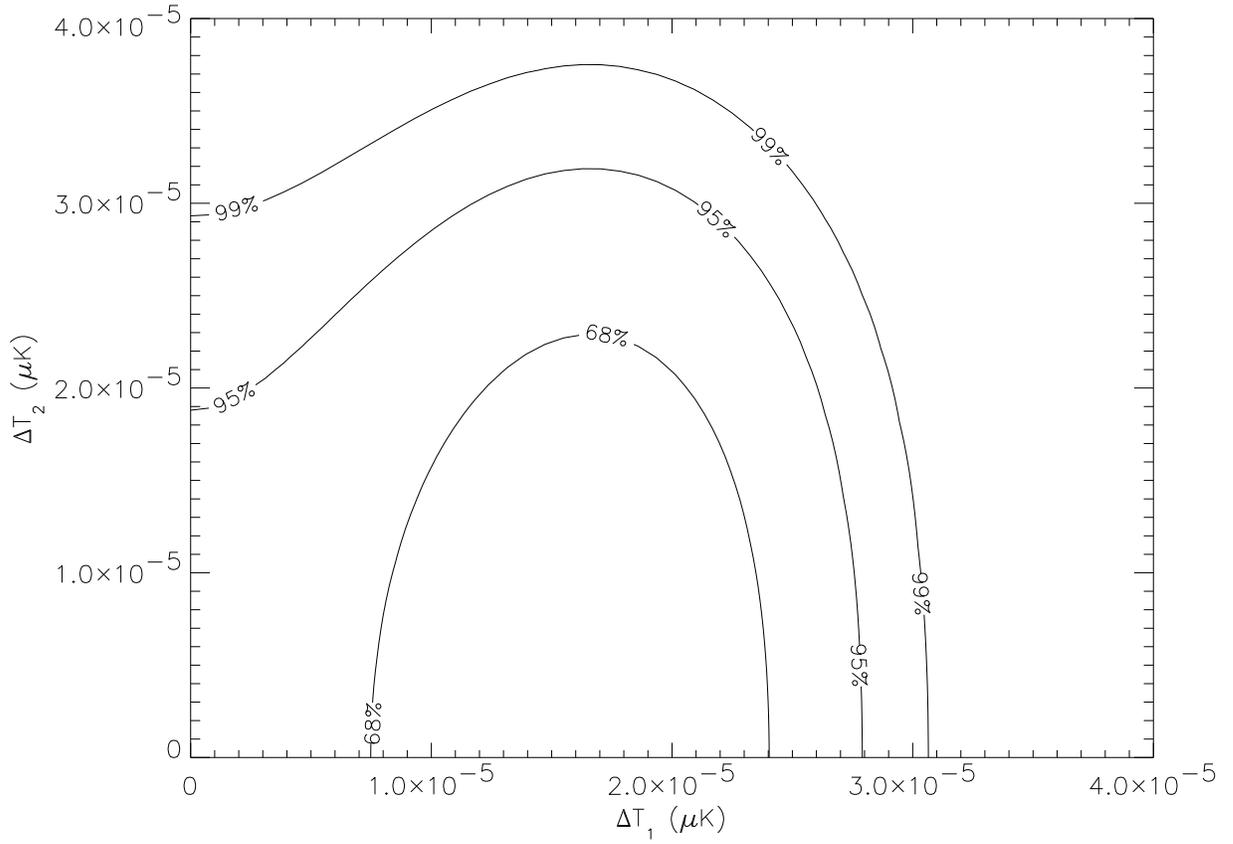}
\figcaption[f6.eps]
{The likelihood that the observed signal in the ten combined fields
is described by flat band power with amplitude $\Delta T_1,\,\Delta T_2$ (measured in $\mu$K).
Contours represent confidence intervals.
}
\label{fig:all2}
\end{figure}

\begin{figure}
\figurenum{7}
\plotone{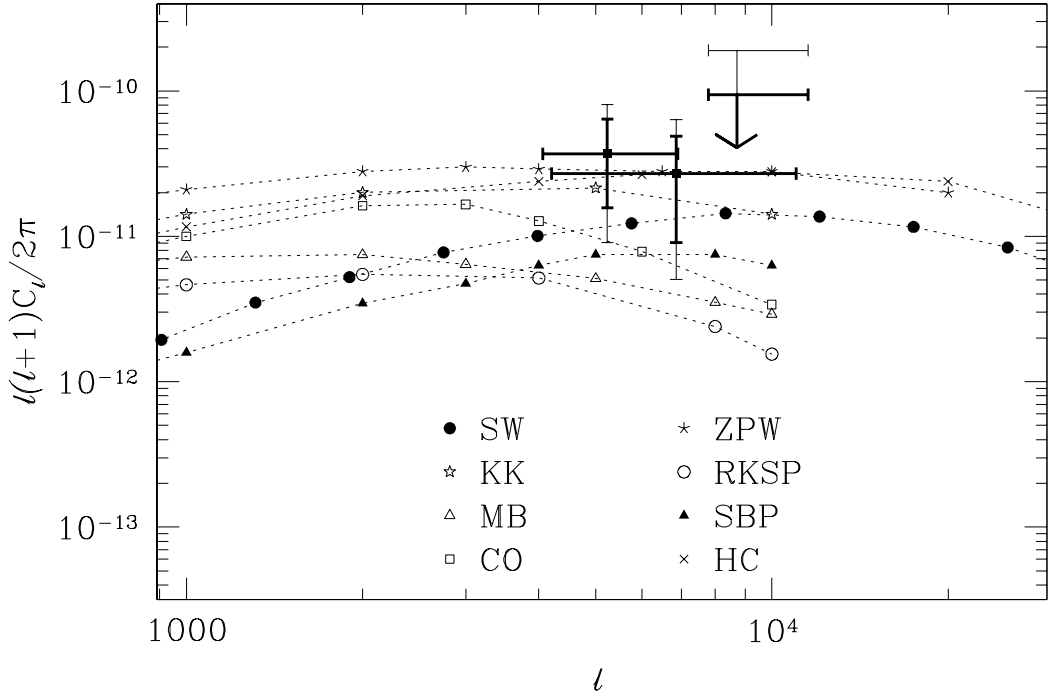}
\figcaption[f7.eps]
{Published estimates of the thermal SZE power
spectrum compared to the result of this work.
The results of the BIMA experiment are expressed with $68\%$ confidence intervals for $\ell_{eff} = 5237$
and for $\ell_{eff} = 6864$ and a $95\%$ upper limit for $\ell_{eff} = 8748$.
Extended error bars represent the estimated effect of non-Gaussian sample variance. 
Other symbols give computations by
the following authors: SW \cite{Vo00}, KK (Komatsu \& Kitayama 1999), MB
(Molnar \& Birkinshaw 2000), CO (Cooray 2000),
ZPW (Zhang et~al. 2002), RKSP (Refregier
et~al.~2000), SBP (Seljak et~al.~2000), and HC
(Holder \& Carlstrom 1999).}
\label{fig:sim}
\end{figure}

\end{document}